\documentclass[aps,prb,amsmath,amssymb,reprint,superscriptaddress]{revtex4-2}

\usepackage{graphicx}
\usepackage{amsmath,amssymb}
\usepackage{xcolor}
\usepackage{color,soul}

\begin{document}

\title{Charge, Bonding, and Optical Properties of the B$_7$Ca$_2$ Cluster: An Alkaline-Earth Dimer Stabilized by a Single Boron Ring}

\author{P. L. Rodr\'iguez-Kessler}
\email{plkessler@cio.mx}
\affiliation{Centro de Investigaciones en \'Optica A.C., Loma del Bosque 115, Lomas del Campestre, Leon, 37150, Guanajuato, Mexico}

\date{\today}

\begin{abstract}
The charge, bonding, and optical properties of the calcium-doped boron cluster B$_7$Ca$_2$ have been systematically investigated using density functional theory calculations. Extensive global basin-hopping searches identify a single-ring B$_7$ geometry stabilized by two calcium atoms symmetrically located on opposite sides of the boron ring as the global minimum. Electronic structure analysis reveals pronounced charge redistribution and strong Ca--B interactions that promote electron delocalization over the boron framework. Hirshfeld charge analysis indicates substantial electron donation from the electropositive calcium atoms to the electron-deficient B$_7$ ring, leading to effective electronic stabilization without the involvement of transition-metal $d$ orbitals. Optical absorption spectra further reflect the delocalized nature of the frontier electronic states. Real-space bonding analyses based on the electron localization function (ELF), Interaction Region Indicator (IRI), and the Laplacian of the electron density reveal a multicenter bonding pattern dominated by electron delocalization within the boron ring, with calcium acting primarily as an electrostatic and charge-donating stabilizer rather than forming localized two-center Ca--B bonds. These results establish B$_7$Ca$_2$ as a prototypical example of an alkaline-earth-metal-stabilized boron ring and highlight the ability of non-transition metals to stabilize aromatic boron clusters through charge transfer and multicenter bonding.
\end{abstract}

\maketitle

\section{Introduction}

Boron clusters have emerged as a central topic in contemporary cluster chemistry because of their exceptional structural flexibility and the predominance of electron-deficient multicenter bonding.\cite{C6CC09570D,C9CP03496J} Unlike conventional covalent systems, boron aggregates readily form delocalized bonding networks that give rise to planar, quasi-planar, and ring-like architectures stabilized by global $\sigma$ and $\pi$ electron delocalization. Small boron clusters, in particular, frequently adopt cyclic or wheel-type motifs that maximize multicenter interactions and electronic stabilization.\cite{doi:10.1021/jp8087918,C6SC02623K}

Introducing metal atoms into boron frameworks has proven to be a powerful approach for accessing new structural and electronic regimes. Metal-doped boron clusters display a wide range of unconventional motifs, including inverse-sandwich structures,\cite{JIA2014128,Zhuan-Yu2014,PHAM2019186,C5CP01650A,LI202325821,RODRIGUEZKESSLER2025117486,b7al2,b7cr2} metal-supported aromatic systems, and highly delocalized ring and wheel-like geometries.\cite{doi:10.1021/acs.inorgchem.7b02585,doi:10.1021/acs.jpclett.0c02656} While the effects of transition-metal dopants have been widely investigated, considerably less attention has been paid to the role of non-transition metals in stabilizing boron clusters.

Alkaline-earth metals such as calcium differ fundamentally from transition metals in that they lack low-lying, partially occupied $d$ orbitals. As a result, their interaction with boron clusters is governed primarily by charge transfer and electrostatic stabilization rather than directional covalent bonding. Electron donation from the metal centers to the electron-deficient boron scaffold can nevertheless induce extensive electron delocalization, providing an alternative stabilization mechanism that does not rely on metal-centered covalency. Recent studies suggest that this charge-transfer-driven effect is sufficient to support planar boron rings and aromatic multicenter bonding in small metal-doped boron clusters.\cite{D5CP01078K,b8cu3,B7Y2,GUEVARAVELA2025115487}

From this perspective, doubly doped systems of the type B$_7$M$_2$ (M = alkaline-earth metal) constitute an attractive yet relatively unexplored class of clusters. In such systems, a small boron ring can be symmetrically stabilized by two electropositive metal atoms, offering an ideal platform to disentangle the roles of electron donation, multicenter bonding, and aromatic stabilization in the absence of transition-metal participation.\cite{B18Ca2,B7Y2}

In the present study, we carry out a detailed computational investigation of the B$_7$Ca$_2$ cluster using extensive global structure searches combined with density functional theory calculations. Although the B$_7$Ca$_2$ cluster has previously been explored in the context of hydrogen storage, its stability, charge distribution, and bonding properties have not been systematically evaluated. Herein, the low-energy structures are characterized through vibrational frequency analysis, atomic charge partitioning, and real-space bonding descriptors derived from the electron density. Our results confirm that the global minimum corresponds to a planar B$_7$ ring stabilized by two calcium atoms located above and below the ring plane. The bonding analysis reveals that calcium primarily functions as an electron donor and electrostatic stabilizer, enabling pronounced $\sigma$ and $\pi$ delocalization within the boron framework rather than forming localized two-center Ca--B bonds. These findings demonstrate the viability of alkaline-earth-metal-stabilized boron motifs as promising building blocks for boron-based functional materials.\cite{C7CP04158F,OLALDELOPEZ2025419}

\section{Computational Details}

Calculations performed in this work are carried out by using density functional theory (DFT) as implemented in the ORCA 6.0.0 code.\cite{10.1063/5.0004608} The exchange and correlation energies are addressed by the PBE0 functional in conjunction with the Def2-TZVP basis set.\cite{10.1063/1.478522,B508541A} Atomic positions are self-consistently relaxed through a Quasi-Newton method employing the BFGS algorithm. The SCF convergence criterion is set to TightSCF in the input file. This results in geometry optimization settings of 1.0e$^{-08}$ Eh for total energy change and 2.5e$^{-11}$ Eh for the one-electron integrals. The  Van  der  Waals  interactions  are  included in the exchange-correlation functionals with empirical dispersion corrections of Grimme DFT-D3(BJ). The electron localization function (ELF) was computed and analyzed using Multiwfn.\cite{https://doi.org/10.1002/jcc.22885} Global minima searches were conducted using standard basin-hopping (BH) algorithm with random rotational–translational perturbations and subsequent DFT local optimization at the PBE0/def2-SVP level.\cite{basin} Low-lying candidates ($<$20 kcal/mol) were reoptimized at PBE0/def2-TZVP level. Vibrational frequency calculations confirmed all minima as true stationary points (no imaginary modes). Spin states from singlet to quintet were evaluated. Spatial region analyses (calc. grid data) were calculated with Multiwfn.

\section{Results and Discussion}

The most stable B$_7$Ca$_2$ cluster adopts a nearly planar single boron ring, stabilized by two calcium atoms positioned above and below the ring plane. This arrangement maximizes electrostatic stabilization through significant charge transfer from Ca to the boron framework. Higher-energy isomers preserve boron flake motifs but display subtle deviations from planarity and variations in Ca coordination, leading to modest energetic penalties. Overall, the structural preferences of B$_7$Ca$_2$ arise from a delicate balance between multicenter boron–boron bonding and predominantly ionic boron–calcium interactions, resulting in a small set of energetically competitive configurations.

\begin{figure}[ht]
\centering
\includegraphics[width=0.45\textwidth]{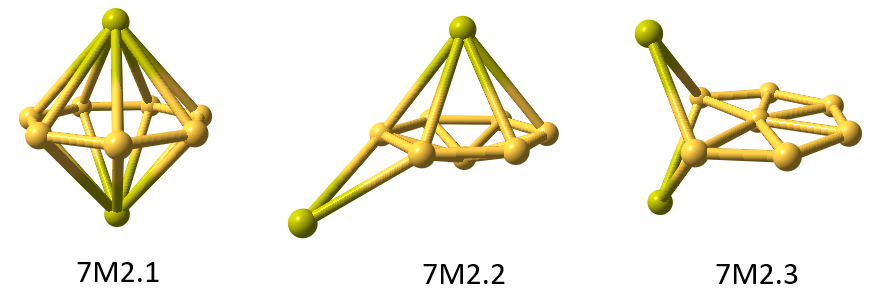}
\caption{Lowest energy structures for B$_{7}$Ca$_2$ cluster at PBE0/def2-TZVP.}
\label{fig_geom}
\end{figure}

The low-energy potential energy surface of B$_7$Ca$_2$ is characterized by several closely lying isomers. Three representative structures are shown in Fig.~\ref{fig_geom}: the single-ring isomer {\bf 7M2.1}, the bent boron flake {\bf 7M2.2}, and the elongated boron flake {\bf 7M2.3}. At the PBE0/def2-TZVP level, {\bf 7M2.2} is marginally more stable than {\bf 7M2.1} by 0.03~eV, indicating a sensitivity of the relative energetics to the choice of functional. To clarify the energetic ordering, additional calculations were performed using the TPSSh/def2-TZVP and $\omega$B97X-D3/def2-TZVP methods. Both approaches identify {\bf 7M2.1} as the global minimum, lying 0.47~eV and 0.38~eV below {\bf 7M2.2}, respectively. These results establish a clear energetic preference for the single-ring structure (see Table~\ref{tab1}).

\begin{table}[h]
\centering
\setlength{\tabcolsep}{6pt} 
\caption{Relative energies of the lowest B$_{7}$Ca$_2$ isomers (in eV). The global minimum is denoted by GM.}
\begin{tabular}{lccc}
\hline\hline
Isomer & PBE0 & TPSSh &  $\omega$B97X-D3 \\
\hline
Single Boron Ring (GM) & 0.03 & 0.00 & 0.00 \\
Bent Boron Flake & 0.00 & 0.47 & 0.38 \\
Elongated Boron Flake & 0.92 & 0.93 & 0.74 \\
\hline\hline
\end{tabular}
\label{tab1}
\end{table}

\subsection{Vibrational and Optical Properties}

\begin{figure}[ht]
\centering
\begin{tabular}{c}
 \resizebox*{0.40\textwidth}{!}{\includegraphics{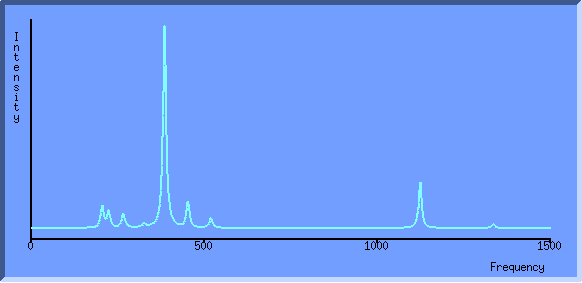}} \\
\end{tabular}
\caption{IR spectra of B$_{7}$Ca$_2$ cluster.}
\label{fig_ir}
\end{figure}

The calculated IR spectrum of the B$_7$Ca$_2$ cluster spans a broad frequency range from the low-frequency region around 140~cm$^{-1}$ up to about 1335~cm$^{-1}$, reflecting the coexistence of soft, collective motions involving the heavy Ca atoms and stiffer vibrations dominated by the covalently bonded boron framework (Fig.~\ref{fig_ir}). Overall, the spectrum is dominated by three main IR-active peaks, located at 376.63, 450.77, and 948.28~cm$^{-1}$, which together capture the essential dynamical behavior of the cluster. Lower-frequency modes below $\sim$300~cm$^{-1}$ are generally weak to moderately intense and are associated with global translations, tilting, and gentle distortions of the B$_7$ ring relative to the Ca atoms. In contrast, the mid- and high-frequency regions are governed by internal deformations of the boron ring that produce much stronger changes in the dipole moment and, consequently, higher IR intensities.

\begin{figure}[ht]
\centering
\begin{tabular}{c}
\resizebox*{0.39\textwidth}{!}{\includegraphics{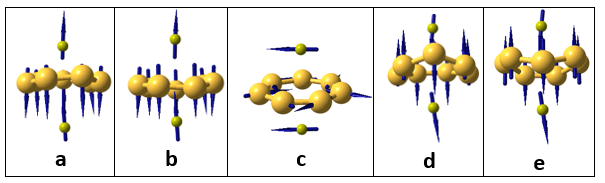}} \\
\end{tabular}
\caption{Representative IR-active vibrational modes of the B$_7$Ca$_2$ cluster.}
\label{fig_modes}
\end{figure}

The most intense feature appears at 376.63~cm$^{-1}$ ($T^2 = 157.01$), which corresponds to a concerted out-of-plane displacement of the B$_7$ ring, as illustrated in panels~{\bf a} and~{\bf b} of Fig.~\ref{fig_modes}. In this mode, the boron framework moves coherently up and down relative to the Ca atoms, resembling a rigid, piston-like oscillation along the principal axis of the cluster. The dominance of this band arises from the large modulation of the dipole moment during this vertical motion, consistent with the strong directional component of the transition dipole.

A second prominent contribution is found in the intermediate-frequency region at 450.77~cm$^{-1}$ ($T^2 = 18.61$), shown in panels~{\bf d} and~{\bf e}. This mode involves a ``butterfly-like'' motion of the B$_7$ ring, where opposite segments bend alternately above and below the mean plane, leading to a dynamic puckering of the ring. Although less intense than the 376.63~cm$^{-1}$ mode, this vibration still produces a substantial IR response due to the pronounced redistribution of charge during the out-of-plane deformation.

The third major IR-active peak occurs at 948.28~cm$^{-1}$ ($T^2 = 35.81$), illustrated in panel~{\bf c}, and corresponds to an in-plane, breathing-like vibration of the boron ring. In this higher-frequency mode, the B--B framework undergoes a nearly symmetric expansion and contraction within the plane, reflecting the intrinsic stiffness of the boron network. The appreciable IR intensity of this band indicates effective coupling between the in-plane ring deformation and the overall dipole moment, likely mediated by charge-transfer interactions with the calcium atoms. Together, these three modes dominate the IR spectrum and highlight the key roles of collective ring motion and B--Ca coupling in shaping the vibrational response of the B$_7$Ca$_2$ cluster.

\begin{figure}[h!]
    \centering
    \includegraphics[width=0.48\textwidth]{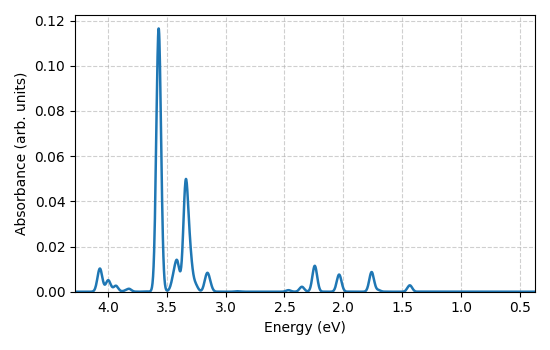}
    \caption{Simulated UV--Vis absorption spectrum of the B$_7$Ca$_2$ cluster calculated at the TD-DFT/TDA level. The spectrum is obtained from the excitation energies and oscillator strengths, applying Gaussian broadening ($\sigma = 0.05$~eV) to simulate a continuous absorption profile. }
    \label{fig:uvvis_b7ca2}
\end{figure}

The UV–Vis absorption spectrum of the B$_7$Ca$_2$ cluster was computed using TD-DFT approximation and exhibits optically allowed transitions extending from the near-infrared to the near-ultraviolet region. The lowest-energy absorption with non-negligible intensity appears at 1.43 eV (State 2, f = 0.0028), followed by several weak-to-moderate transitions throughout the visible region. A more pronounced feature is observed at 2.24 eV (State 15, f = 0.011), marking the onset of significant optical activity. In the near-UV region, several intense absorptions dominate the spectrum, notably at 3.31 eV (State 37, f = 0.013), 3.34 eV (State 40, f = 0.045), 3.41 eV (State 41, f = 0.012), and 3.57 eV (State 47), the latter being the most intense transition with an oscillator strength of f = 0.117. These high-energy excitations account for the main peaks shown in Fig.~\ref{fig:uvvis_b7ca2} and reflect strongly allowed electronic transitions associated with the delocalized boron framework, enhanced by charge donation from the calcium atoms. Overall, the absorption profile confirms that B$_7$Ca$_2$ is optically active across a broad spectral window, with its dominant features governed by collective excitations of the boron ring stabilized by alkaline-earth metal doping.

\subsection{Charge analysis}

\begin{table}[h!]
\centering
\caption{Atomic charges of the B$_7$Ca$_2$ cluster obtained from standard Hirshfeld and atomic dipole-corrected Hirshfeld (ADCH) analyses. All values in $e^-$.}
\begin{tabular}{c c c}
\hline
Atom & Hirshfeld charge & ADCH charge \\
\hline
Ca1 & +0.6485 & +0.9860 \\
Ca2 & +0.6476 & +0.9856 \\
B3  & -0.1679 & -0.2681 \\
B4  & -0.1693 & -0.2695 \\
B5  & -0.1867 & -0.2883 \\
B6  & -0.2100 & -0.3022 \\
B7  & -0.2116 & -0.3034 \\
B8  & -0.1836 & -0.2874 \\
B9  & -0.1638 & -0.2526 \\
\hline
\end{tabular}
\label{tab:charges_B7Ca2}
\end{table}

\begin{figure*}[ht!]
\centering
\begin{tabular}{cc}
\resizebox*{0.32\textwidth}{!}{\includegraphics{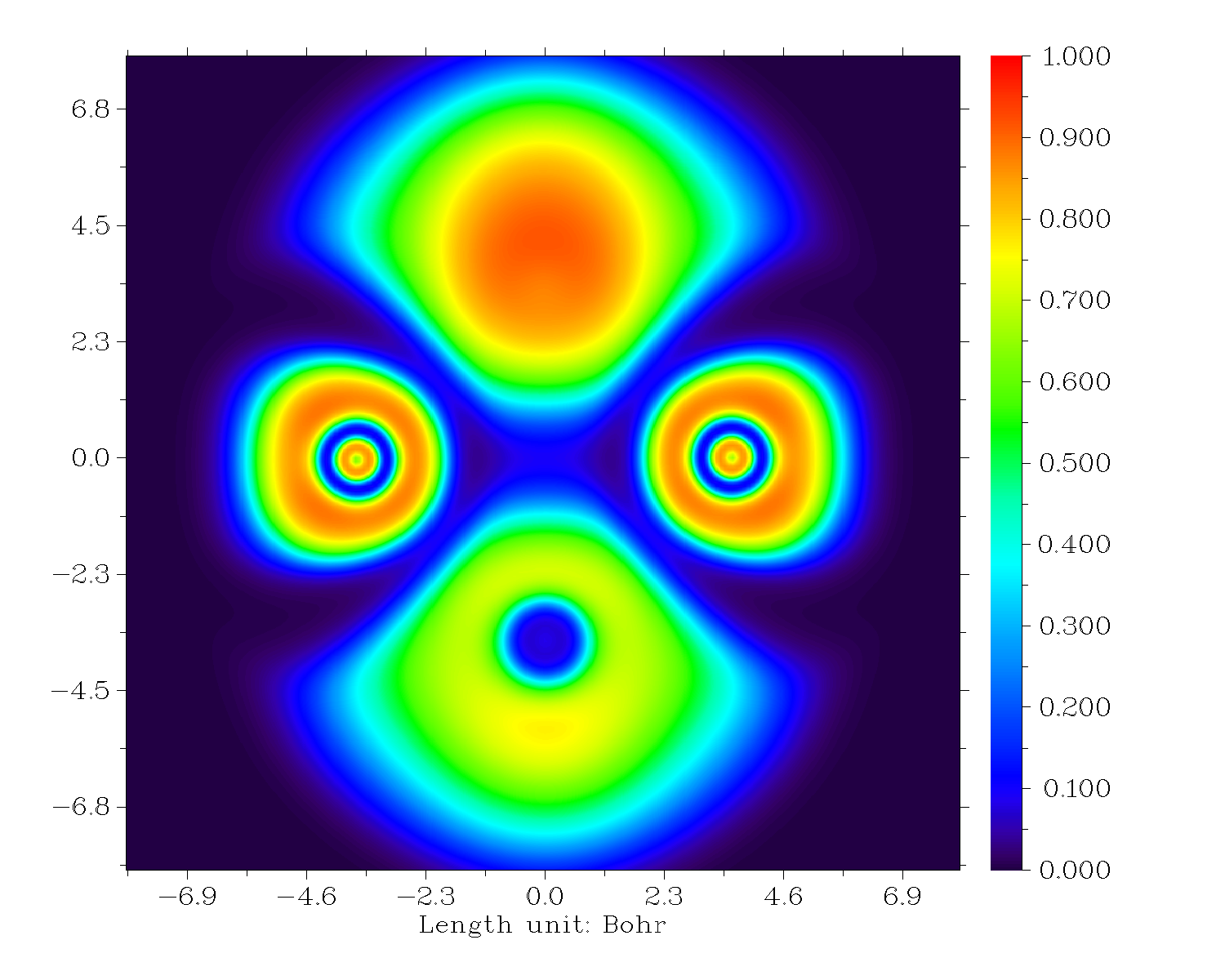}} &
\resizebox*{0.32\textwidth}{!}{\includegraphics{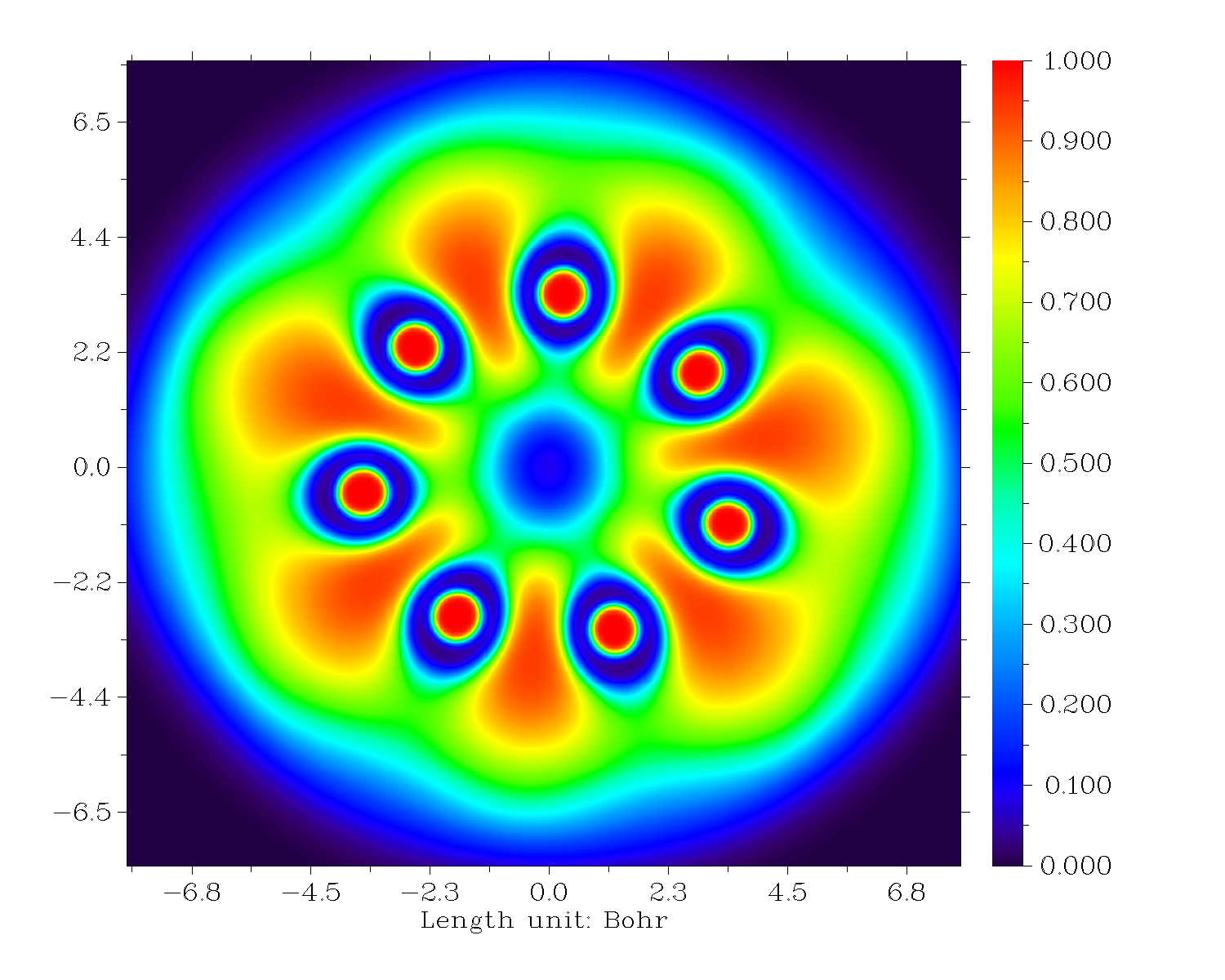}}
\end{tabular}
\caption{ELF 2D maps of the B$_{7}$Ca$_2$ cluster. Left: side view showing localized Ca-centered ELF maxima. Right: top view showing a continuous $\sigma$-delocalized network along the boron single ring.}
\label{fig_elf}
\end{figure*}

To obtain a more accurate description of charge transfer in the B$_{7}$Ca$_2$ cluster, atomic dipole–corrected Hirshfeld (ADCH) charges were evaluated using Multiwfn.\cite{Lu2012} The ADCH method accounts for intra-atomic polarization effects, providing a more reliable estimate of charge separation than conventional Hirshfeld analysis. As summarized in Table~\ref{tab:charges_B7Ca2}, the results reveal a pronounced electron transfer from the calcium dopants to the boron framework, with each Ca atom carrying a positive charge of approximately +0.986 $e^-$. Correspondingly, the excess electronic density is distributed over the seven boron atoms, which acquire negative charges ranging from -0.252 to -0.303 $e^-$. The relatively uniform charges on the boron atoms indicate a delocalized electron distribution within the boron scaffold. The nearly identical charges on the two Ca atoms reflect their symmetric placement above and below the boron framework and confirm their equivalent electronic roles. Overall, the ADCH charge analysis supports a bonding picture in which calcium acts as an efficient electron donor, stabilizing the electron-deficient boron cluster through predominantly ionic B–Ca interactions, while the donated electrons are extensively delocalized over the boron atoms. This analysis highlights the critical role of calcium-induced polarization in reinforcing the stability of the B$_{7}$Ca$_2$ motif.

\subsection{Electron Density–Based Bonding Analysis}

\begin{figure}[h]
\centering
\begin{tabular}{c}
\resizebox*{0.42\textwidth}{!}{\includegraphics{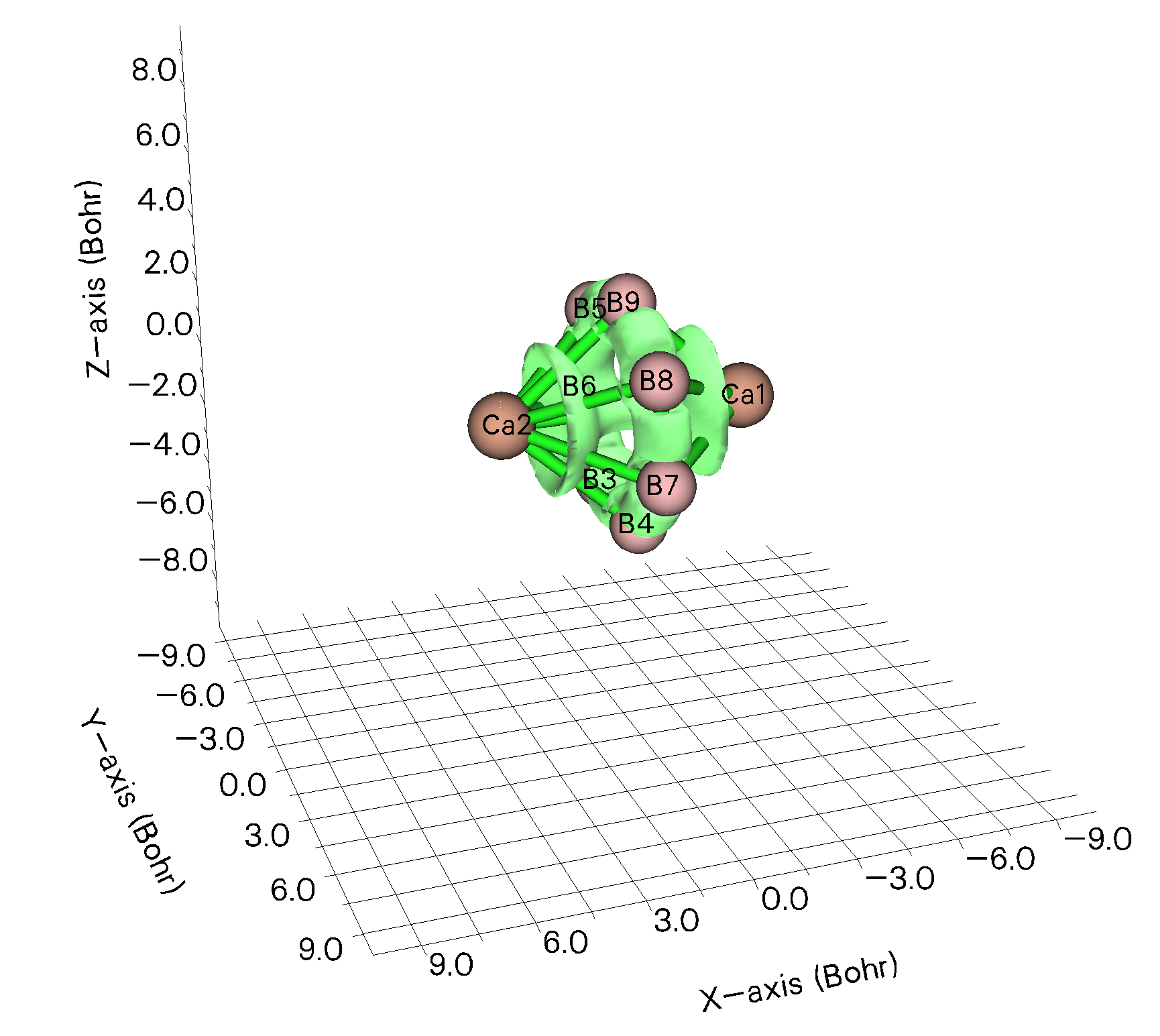}} \\
\resizebox*{0.42\textwidth}{!}{\includegraphics{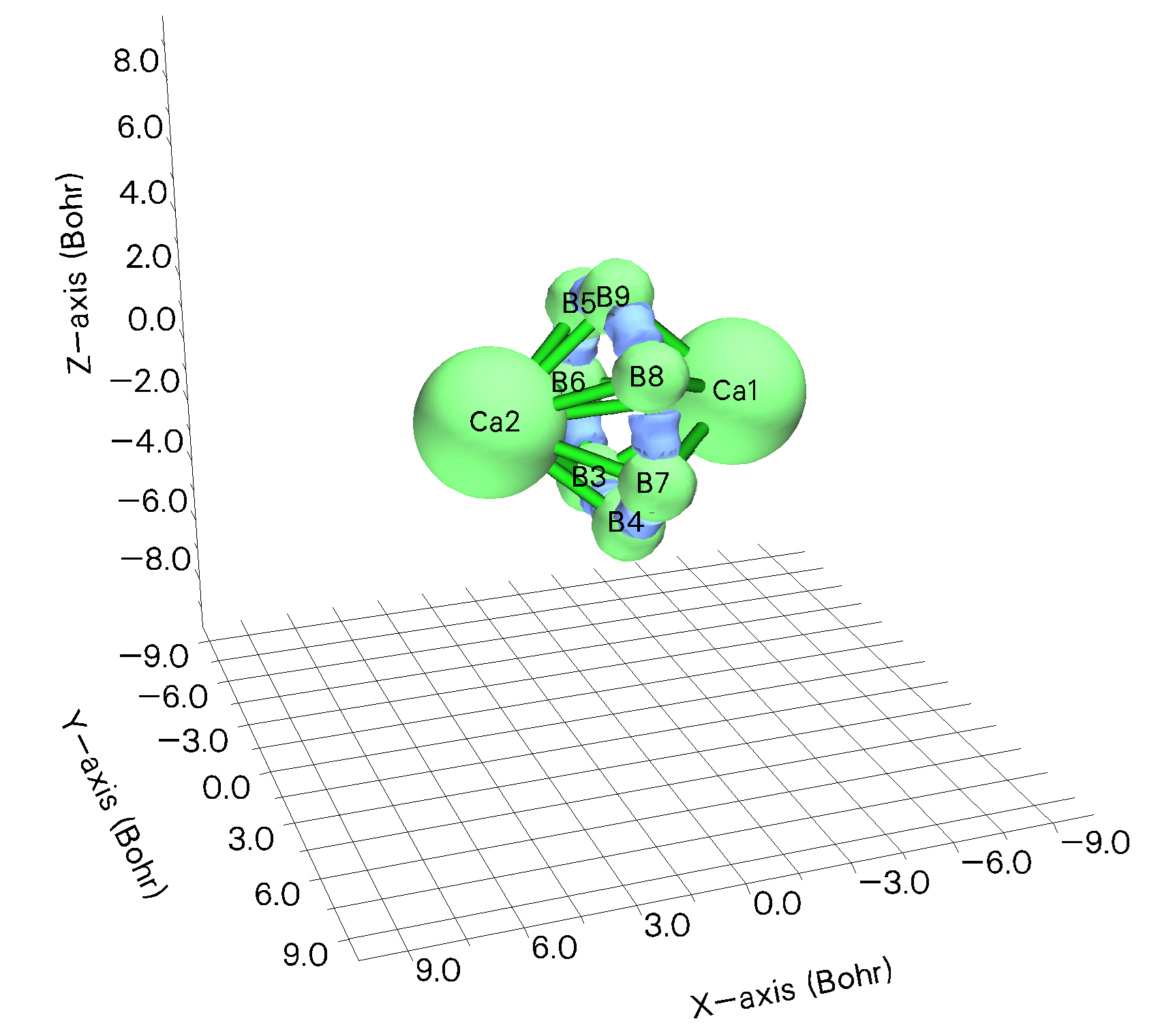}} \\
\end{tabular}
\caption{Laplacian of electron density ($\nabla^2\rho$) and Interaction region indicator (IRI) of B$_{7}$Ca$_2$ clusters.}
\label{fig_elf_lol}
\end{figure}

The ELF maps of B$_7$Ca$_2$ reveal clear electron localization patterns that illustrate the interaction between the calcium dopants and the boron ring (Fig.~\ref{fig_elf}). In the side view (left panel), the ELF exhibits strong maxima centered on each Ca atom, highlighting significant Ca–B interactions and electron donation from calcium to boron. Surrounding these maxima are smaller ELF lobes indicative of multicenter Ca–B–B bonding, which stabilizes the calcium atoms along the boron ring. In the top view (right panel), the ELF forms a nearly continuous annular pattern along the 7-member boron ring, demonstrating a delocalized $\sigma$-network around the single ring. The uniform and circular ELF distribution confirms the presence of global delocalization and aromatic-like electron circulation, while the central cavity remains electron-deficient, indicating the absence of inner-core bonding. Overall, these ELF features suggest that the stability of B$_7$Ca$_2$ arises from a combination of Ca-mediated charge transfer and extended $\sigma$-delocalization over the boron single ring.

The Laplacian of the electron density ($\nabla^2\rho$) for B$_{7}$Ca$_2$ highlights the characteristic distribution of charge concentration and depletion that underpins its multicenter bonding framework. Regions of negative $\nabla^2\rho$ around the B–B contacts indicate localized charge concentration associated with the multi-centered $\sigma$-bonds within each boron ring, whereas the predominantly positive Laplacian surrounding the Ca centers reflects charge depletion consistent with substantial Ca→B electron donation rather than covalent Ca–B bonding. This pattern corroborates the electron-transfer–driven stabilization mechanism inferred from ELF analysis. Complementarily, the Interaction Region Indicator (IRI) map reveals a continuous green toroidal belt encircling the boron framework, characteristic of extended, noncovalent but strongly attractive delocalized interactions that unify the two rings into a coherent $\sigma$-aromatic manifold. The absence of sharp red or blue IRI features between the Ca atoms and the boron cage further confirms that Ca participates through diffuse, multicenter electrostatic interactions rather than localized two-center bonds. 

\section{Conclusions}

In conclusion, a comprehensive density functional theory investigation has been performed to elucidate the structural, electronic, bonding, and optical properties of the calcium-doped boron cluster B$_7$Ca$_2$. Extensive global basin-hopping searches identify a planar single-ring B$_7$ framework symmetrically stabilized by two calcium atoms as the global minimum, confirming its pronounced energetic and dynamical stability. Electronic structure and Hirshfeld charge analyses reveal substantial electron donation from the electropositive calcium atoms to the electron-deficient boron ring, promoting strong charge redistribution and pronounced $\sigma$ and $\pi$ electron delocalization without the involvement of transition-metal $d$ orbitals. Real-space bonding analyses show that the stabilization of B$_7$Ca$_2$ is dominated by multicenter delocalized bonding within the boron ring, with calcium acting primarily as an electrostatic and charge-donating stabilizer rather than forming localized two-center Ca--B bonds. The calculated optical absorption spectrum further reflects the delocalized nature of the frontier electronic states. Overall, these results confirm the robust structural and electronic stability of B$_7$Ca$_2$, supporting its viability as a promising building block for prospective applications in boron-based functional materials.


\section{Acknowledgments}
P.L.R.-K. would like to thank the support of CIMAT Supercomputing Laboratories of Guanajuato and Puerto Interior. The authors thankfully acknowledge the computer resources, technical expertise and support provided by the Laboratorio Nacional de Supercómputo del Sureste de México, CONACYT member of the network of national laboratories.



\bibliographystyle{unsrt}
\bibliography{mendelei.bib}
\end{document}